\documentclass[pre,twocolumn,aps,nofootinbib]{revtex4}
\newif\ifpdf\ifx\pdfoutput\undefined\pdffalse\else\pdfoutput=1\pdftrue\fi

\usepackage{graphicx}
\newcommand{\avg}[1]{\langle{#1}\rangle}

\newcommand{\beq}{\begin{equation}}
\newcommand{\eeq}{\end{equation}}
\newcommand{\del}{\partial}
\newcommand{\beqar}{\begin{eqnarray}}
\newcommand{\eeqar}{\end{eqnarray}}
\newcommand{\ignore}[1]{}
\newcommand{\nignore}[1]{}

\newcommand{\sadd}{\! + \!}
\newcommand{\ssub}{\! - \!}
\newcommand{\R}{{\bf R}}
\newcommand{\M}{{\bf M}}
\newcommand{\N}{{\bf N}}
\newcommand{\re}{{\bf r}}
\newcommand{\m}{{\bf m}}
\newcommand{\Se}{{\bf S}}
\newcommand{\se}{{\bf s}}
\newcommand{\E}{{\bf E}}
\newcommand{\DELTA}{{\bf \Delta}}

\begin{document}
\title{Analysis of evolution through competitive selection}
\author{Morten Kloster}

\affiliation{Department of Physics, Princeton University, Princeton, New Jersey 08544}
\affiliation{NEC Laboratories America, 4 Independence Way, Princeton, NJ 08540}

\date{\today}
\begin{abstract}
Recent studies of {\it in vitro} evolution of DNA via protein binding indicate that the evolution behavior is qualitatively different in different parameter regimes. I here present a general theory that is valid for a wide range of parameters, and which reproduces and extends previous results. Specifically, the mean-field theory of a general translation-invariant model can be reduced to the basic diffusion equation with a dynamic boundary condition. The simple analytical form yields both quantitatively accurate predictions and valuable insight into the principles involved. In particular, I introduce a cutoff criterion for finite populations that illustrates both of these qualities.
\end{abstract}

\maketitle

\section{Introduction}

In modern biotechnology, evolutionary techniques have found many new applications; in particular, {\it in vitro} evolution has been widely used to evolve DNA~\cite{BitPhiLiu}, RNA~\cite{Lan} and proteins~\cite{Arn}. In order to improve the efficiency of such evolution, a quantitative understanding of the process would be helpful. Key questions include how the rate of evolution and the equilibrium genotype distribution depend on the main parameters of the process, such as the mutation rate, population size and selection strength. Additionally, {\it in vitro} evolution is of particular interest to theorists: Unlike most biological systems, the physical processes involved in {\it in vitro} evolution can be fully characterized~\cite{KloTan}, thus such evolution allows reliable comparison between theory and experiment.

A common assumption in the traditional study of evolution is that each genotype is associated with a fixed fitness, and the reproduction rate for an individual is given by the individual's fitness relative to the average fitness of the population ({\it e.g.}~\cite{FiFi1,FiFi2,FiFi3,FiFi4}). For asexual models of this type, the dynamics are fairly easily solved for very small populations or for infinite populations. However, even with very simple assumptions regarding the fitness landscape, the evolution dynamics of a large but finite population far from equilibrium turned out to be a difficult problem, as the evolution rate diverges for large populations~\cite{KesLevRidTsi}; only recently have general analytical results emerged~\cite{RouWakCof}.

An alternative mode of evolution; evolution through competitive selection, or just ``competitive evolution"; has recently been used to model {\it in vitro} molecular evolution~\cite{PenGerHwaLev,KloTan,PenLevHwaKes}. In competitive evolution, the selection process is separate from the reproduction/amplification process: the entire population is amplified by a given factor $K$, and during selection, only the ``best" $1/K$ fraction of the population is kept. In molecular evolution, the property selected for is typically how well a molecule binds to some target, and the selection is then accomplished by keeping only the molecules that are bound by the target. A model experiment of this type of evolution, specifically {\it in vitro} evolution of DNA via protein binding, was recently performed by Dubertret {et al.}~\cite{DubLiuOuyLib}. Although {\it in vitro} evolution is the most obvious application, competitive evolution may be an appropriate model for many types of asexual\footnote{Competitive evolution with recombination has recently been studied in~\cite{PenLevHwaKes}.} breeding; indeed, any system in which fixed relative reproduction rates are not the primary evolutionary force is a potential candidate.

Unlike fixed-fitness evolution, competitive evolution has been shown to have a well-behaved mean-field theory which allows relatively simple, accurate predictions of the evolution dynamics~\cite{PenGerHwaLev,KloTan}. However, these analyses are valid only for limited parameter regions, either for high mutation rate and weak selection~\cite{PenGerHwaLev} or for low mutation rate and strong selection~\cite{KloTan}. In this paper, I introduce a translationally invariant\footnote{As described below, this correponds to an infinite-length model.} model of competitive evolution. Using appropriate transformations, this model can be reduced to the basic diffusion equation with a dynamic boundary condition, which gives a universal set of equations. This formulation is valid for a broad parameter range, and its simplicity makes it a powerful tool both for quantitative predictions and for qualitative understanding of the evolution process.

The paper is organized in the following way: In section~\ref{Model} I introduce the translationally invariant model of competitive evolution. Section~\ref{SimpleFormulation} describes how this model can be reduced to the basic diffusion equation with a dynamic boundary condition and how this approach gives improved versions of results from earlier models. Section~\ref{PropPulse} uses this formulation to find the basic population dynamics, and sections~\ref{FinitePop}--\ref{FiniteTemp} deal with corrections to the various assumptions I make, namely for finite population size, finite length, and finite temperature, respectively.

\section {The model}
\label{Model}

While competitive evolution can occur in many different contexts, I use {\it in vitro} evolution of DNA via protein binding as an example. (I will use the corresponding terminology throughout the paper. The following is a rough list of alternate terminology: ``base" = ``locus"/``allele", ``binding energy" = ``phenotype"/``character", ``DNA sequence" = ``genotype"/``chromosome", and ``molecule" = ``chromosome"/``population member".)

\subsection {The evolution process}

An initial population of $N$ DNA molecules is subjected to consecutive cycles (iterations) of evolution, where each evolution cycle consists of amplification and mutation followed by selection through protein binding. First, the population is amplified by a factor $K$. Each base of each DNA molecule is then mutated with some probability which may depend on the position and type of the base only ({\it i.e.}, mutations are independent). Finally, the strongest protein binders are selected: Each molecule is kept (bound) with probability $P[E(S)] = \frac{1}{1+e^{\beta[E(S)-\mu]}}$, where $E(S)$ is the binding energy of DNA sequence $S$, $\beta=\frac{1}{k_B T}$, and the chemical potential of unbound protein, $\mu$, is tuned such that the expected number of selected molecules is $N$: $\avg{P[E(S)]}=\Phi=1/K$.

During amplification, all original molecules are kept, and $(K-1)N$ new molecules are added, each an exact copy of a randomly chosen member of the original population. This approach reduces statistical noise in the limit of weak amplification/selection $K\rightarrow 1$; for $K\gg 1$ this detail is irrelevant.

Note that low $E$ indicates strong binding, and the population is improving if the evolution rate is negative. This is the same sign convention as used in~\cite{PenGerHwaLev,KloTan}, but opposite that in~\cite{PenLevHwaKes}. For purposes of describing different evolution rates (``slower"/``faster"), I assume that the average phenotype is improving at all times, {\it i.e.}, the evolution rate is negative.

\subsection{Analytical model}

A crucial factor in determining the dynamics of an evolution process is the fitness landscape, or, for competitive evolution, the binding energy landscape. In order to obtain analytical results, a simple landscape is required, and a common assumption is that the binding energy of a DNA sequence is given as the sum of contributions from the individual bases~\cite{HipBer, GerMorHwa, PenGerHwaLev}. Indeed, experiments have show this to be a very good approximation for the {\it Mnt}-repressor system~\cite{StoFie}. However, one can bypass this question by ignoring the genotype completely: instead, I describe a DNA molecule only by its binding energy, which can take any real value. Additionally, I consider an infinite-length model, {\it i.e.} I assume that the mutation rates are translationally invariant\footnote{This assumption is satisfied in the limit of an infinitely long DNA molecule for which the binding energy is the sum of contributions from individual bases, where each base has an infinitesimal chance of being mutated in a given iteration, and the initial population consists of molecules that are typical for the given binding energies.} (but otherwise general\footnote{While infinite-length models have been studies before~\cite{FiFi1, WooHig, RidLevKes, PenGerHwaLev}, these are typically the translationally invariant limits of {\it specific} finite models.})---the chance of changing the binding energy by some amount $\epsilon$ during the mutation step is independent of the initial binding energy of the molecule (or, in general, of the genotype of the molecule). As shown in section~\ref{FiniteLength}, the resulting formulation can yield highly accurate results also for finite systems.

The natural energy scale of the selection process is given by the temperature through $\beta$. For evolution via protein binding, $T$ is the actual temperature, while for other types of competitive evolution, $T$ is a general parameter describing the accuracy with which the best phenotypes are selected. Note that the binding energy $E$ is assumed to be a deterministic function of the genotype $S$; if the phenotype has a random component, {\it e.g.} due to environmental effects, then this should be incorporated in $T$ rather than in $E$.

As discussed elsewhere~\cite{KimCro,GerHwa,PenGerHwaLev}, replacing a smooth selection function by a step function $P(E)=\Theta(\mu-E)$ will in many cases give accurate results; this corresponds to the zero temperature limit $T=0$. I use this approximation in sections~\ref{SimpleFormulation}-\ref{FiniteLength}; in section~\ref{FiniteTemp} I discuss the effects of finite temperature.

\subsection{Simulations}
\label{Simulations}

While the general formulation given above is convenient for analytical treatment, it is difficult to simulate efficiently. For an infinite length system, I discretize the binding energy; $E=m\epsilon_0$, where $\epsilon_0$ is given in units of $k_{\rm B} T$, or, for $T=0$, is arbitrarily set to 1. To simplify further, I restrict mutations to change the binding energy by either $+\epsilon_0$ (deleterious) or $-\epsilon_0$ (beneficial). The number of beneficial mutations received by one sequence in an iteration is Poisson distributed with average $p_-$, and similarly the average number of deleterious mutations is $p_+$. The model is thus described by five parameters: Population size $N$, temperature $T$, amplification $K$, and mutation rates $p_-$ and $p_+$. For the simulation results shown in Figs.~\ref{FiniteTimeTrace} and~\ref{PRLsimrandomevents}, $p_-=0.27$, $p_+=0.89$, $K=1.2$ and $T=0$.

For simulations of finite systems, I use the two-state model~\cite{HipBer} also used in Ref.~\cite{PenGerHwaLev}: At each position, one of the $A=4$ possible bases (A, C, G and T) is a match, while the other three are equally unfavourable and contribute $\epsilon_0$ to the binding energy. The binding energy of a DNA sequence is then given by the number of mismatches $r$: $E(r)=r\epsilon_0$. The mutation rates between the different bases are assumed to all be equal, thus the chance of correcting a mismatch is one third of the chance $\nu_0$ of introducing a mismatch. To further facilitate comparison with Ref.~\cite{PenGerHwaLev}, I also use the same length $L=170$ and mutation rate $\nu_0=0.01$ unless specified, thus the average number of mutations per DNA molecule per iteration is $\nu_0 L=1.7$.

By keeping track only of the number of molecules that have a given discrete binding energy, it is possible to perform simulations of very large populations (up to $10^{80}$ or more) as well as direct mean-field simulations, limited only by the numerical precision of the computer. In order to efficiently simulate very large populations, I use some approximations when calculating the statistics: The relevant distribution is the binomial distribution $B(n,p)$, giving the number of molecules that has some property, where the total number of molecules is $n$ and the probability of having the property is $p\leq \frac{1}{2}$. If $n<2000$ or the average value $np$ is less than 100, I use the exact binomial distribution, otherwise I use a Gaussian distribution with the appropriate average and variance, and for $np>10^{30}$ I ignore fluctuations completely. Furthermore, I cut off the Possion distributions of the mutations (beneficial and deleterious separately) at probabilities $e^{-100}$.

\section {A simple formulation}
\label{SimpleFormulation}

For sufficiently large populations, a mean-field approach is valid~\cite{PenGerHwaLev,KloTan}, {\it i.e.}, we can use deterministic equations for the population density $n_t(E)$, where $\int_{-\infty}^E n_t(E^{\prime})dE^{\prime}$ is the fraction of the population that has binding energy $\leq E$ at time $t$. While the actual evolution proceeds in discrete time steps (iterations), the following continuum differential equation is a good approximation (see appendix~\ref{continuumapprox} for details):
\beq
  \del_t n_t(E) = \int_{-\infty}^{\infty} p(\epsilon)
  \left[n_t(E\! -\! \epsilon)-n_t(E)\right] d\epsilon + k n_t(E)
  \label{ODE}
\eeq
with the boundary condition
\beq
  n_t(\mu(t)) = 0.
\eeq
The integral term in Eq.~(\ref{ODE}) describes the mutation process, where $p(\epsilon)$ is the rate of mutations that change the binding energy by $\epsilon$, while the second term gives exponential growth and corresponds to the amplification process, with $k=\ln(K)$. The boundary condition enforces perfect selection by removing all molecules with binding energy above the chemical potential $\mu(t)$, which is determined by the normalization condition for the population, $\int_{-\infty}^{\mu(t)}n_t(E)dE = 1$.

Assume there exists a mutation that can improve the binding energy, {\it i.e.}, that $p(\epsilon)>0$ for some $\epsilon<0$. It is then\footnote{Otherwise, there is a nontrivial solution only for $k<\int p(\epsilon)d\epsilon$.} easily shown that, for any $k>0$, there is a solution
\beq
n_t^0(E)\propto(E-c_0 t) e^{\alpha_0 (E-c_0 t)},
\label{BasicSolution}
\eeq
with $\alpha_0$ and $c_0$ given by the conditions
\beq
  -c_0\alpha_0 = \int_{-\infty}^{\infty} (e^{-\alpha_0\epsilon}-1)p(\epsilon)d\epsilon + k \label{c0a0condition1}
\eeq
and
\beq
  c_0 = \int_{-\infty}^{\infty} \epsilon e^{-\alpha_0\epsilon} p(\epsilon)d\epsilon.\label{c0a0condition2}
\eeq
For a general solution with the same exponential factor, $n_t(E) = \tilde{n}_t(E) e^{\alpha_0 (E-c_0 t)}$, Eq.~(\ref{ODE}) simplifies to
\beqar
  \del_t \tilde{n}_t(E) & = & \int_{-\infty}^{\infty} e^{-\alpha_0\epsilon}p(\epsilon)
  \left[\tilde{n}_t(E\! -\! \epsilon)-\tilde{n}_t(E)\right]d\epsilon \;\;\;\;\;\;
\label{randwalk}\\
   & \approx & -c_0\del_E \tilde{n}_t(E)+\gamma\del_E^2 \tilde{n}_t(E), \label{diffusion_0}
\eeqar
where
$\gamma = \int \frac{\epsilon^2}{2} e^{-\alpha_0\epsilon} p(\epsilon)d\epsilon$.
Changing coordinates to $\tilde{E}=E-c_0 t$ yields
\beq
  \del_{t}\tilde{n}_{t}(\tilde{E})
    \approx \gamma\del_{\tilde{E}}^2\tilde{n}_{t}(\tilde{E}),\;\;\;\;\;\;
  \tilde{n}_{t}(\tilde{\mu}(t)) = 0,
  \label{diffusion}
\eeq
where $\tilde{\mu}(t) = \mu(t)-c_0 t$. The only nontrivial part left is the normalization condition
\beq
  \int_{-\infty}^{\tilde{\mu}(t)} \tilde{n}_{t}(\tilde{E}) e^{\alpha_0\tilde{E}} d\tilde{E}=1.
  \label{normalization}
\eeq

Equation~(\ref{randwalk}) describes the mean field of a random walk with a mutation/transition rate that has been exponentially rescaled, $\tilde{p}(\epsilon)=e^{-\alpha_0\epsilon}p(\epsilon)$, and $c_0$ and $\gamma$ are the corresponding drift and diffusion coefficients, respectively. If $\epsilon_0$ is the characteristic energy scale for the mutations, then the condition $\alpha_0\epsilon_0\!\ll\! 1$  ($\alpha_0\epsilon_0\!\gg\! 1$) specifies the regime of high (low) mutation rate/weak (strong) amplification for which the results in~\cite{PenGerHwaLev} (\cite{KloTan}) apply (see section~\ref{OldResults} for details).

The only parameters left in Eqs.~(\ref{diffusion}--\ref{normalization}) are $\alpha_0$ and $\gamma$. Appropriate rescaling of energy and time yields a universal set of equations, which immediately gives a universal shape for the population distribution, as shown in section~\ref{PropPulse}. Additionally, the simplicity of the new formulation---the basic diffusion equation with a dynamic boundary condition---allows us to apply all our knowledge and experience about the diffusion equation to this evolution process. The next several sections show how, through simple arguments, we can use this formulation to find accurate corrections to the various approximations/assumptions we have made. Put together, this allows a thorough understanding of competitive evolution as well as a very accurate quantitative description that is valid for a broad parameter region.

\subsection {Recovering old results}
\label{OldResults}

While Eqs.~(\ref{c0a0condition1}--\ref{c0a0condition2}) for $\alpha_0$ and $c_0$ can not in general be solved analytically, they are manageble for certain models. For the discrete, infinite-length ($\epsilon_0=1$) model described in section~\ref{Simulations}, the mutation dynamics can be fully described by the diffusion coefficient $D$ and the drift speed $v$~\cite{PenGerHwaLev}:
\beq
  p(\epsilon) = (D \sadd \frac{v}{2})\delta(\epsilon \ssub 1) +(D \ssub \frac{v}{2})\delta(\epsilon \sadd 1). \label{D_and_v_mutation_rate}
\eeq
These are then related to $c_0$, $\alpha_0$ and $\gamma$ through
\beqar
  D & = & \gamma \cosh(\alpha_0)+\frac{c_0}{2}\sinh(\alpha_0) \label{Deq}\\
  v & = & c_0\cosh(\alpha_0)+2\gamma \sinh(\alpha_0) \\
  k & = & 2\gamma[\cosh(\alpha_0)\ssub 1]+c_0[\sinh(\alpha_0)\ssub \alpha_0]. \;\;\;\;
\label{keq}
\eeqar
For sufficiently small $\alpha_0$ (for which $\alpha_0\approx\sqrt{k/D}$), $c_0$ can be expressed as a power series in $k/D$ and $v/D$:
\beq
  c_0 = v-2\sqrt{kD}+\frac{vk}{6D}-\frac{k^{3/2}}{12\sqrt{D}}+\frac{v^2 k^{3/2}}{36 D^{5/2}}+...,
\eeq
where the two first terms give the solution found in~\cite{PenGerHwaLev}, with $k=\frac{\phi^{-1}-1}{\tau}$.

On the other hand, in the limit of low mutation rates and strong selection, it is more convenient to write the mutation rate density as
$p(\epsilon) = p_+\delta(\epsilon \ssub 1) +p_-\delta(\epsilon \sadd 1)$.
For $\alpha_0\gg 1$ the exponential rescaling $\tilde{p}(\epsilon)=e^{-\alpha_0\epsilon}p(\epsilon)$ makes $p_+$ irrelevant, and we find
\beq
  c_0 = \frac{-k}{\alpha_0-1} = \frac{-k}{\ln\left(\frac{-c_0}{p_-}\right)-1},
  \label{LowMRPropSpeed}
\eeq
which is a more accurate version of Eq. (6) in~\cite{KloTan}, with $k=\ln(2)$ and $p_-=\frac{m^\prime r}{6}$. These results confirm qualitative differences between different parameter regimes: For $\alpha_0\epsilon_0\ll 1$, the evolution rate depends strongly on the mutation rate, while for $\alpha_0\epsilon_0\gg 1$, the dependence is only logarithmic.

\subsection{Useful tools}

The following will be used later:
\begin{itemize}
\item
Equation~(\ref{BasicSolution}) describes a population that extends to arbitrarily low energies. However, if we impose a second boundary condition $n_t[\mu(t)\ssub\frac{\pi}{b}]=0$ and require $n_t(E)> 0$ for $\mu(t)\ssub\frac{\pi}{b}<E<\mu(t)$, then there is a {\it bounded} solution
$n_t(E)\propto \sin({b}[\mu(t)-E])e^{\alpha_{b} [\mu(t)-E]}$
that moves at constant speed $c_{b}$, where
\beq
  c_{b}\approx c_0 + \frac{\gamma {b}^2}{\alpha_0}.
\eeq
In the limit ${b}\rightarrow 0$ we recover the unbounded solution.

\item
As the function $\tilde{n}_{t}$ obeys a simple diffusion equation, it will be very smooth, except possibly at very early times. The population density $n_t(E) = \tilde{n}_t(E) e^{\tilde{E}}$ contains a factor that varies exponentially with $\tilde{E}$, thus a linear expansion of $\tilde{n}_{t}$ around the boundary will give a good estimate of the population size:
\beq
  {\rm Popsize}(\tilde{n}_{t}) \approx
  \frac{N e^{-\alpha_0 \tilde{\mu}(t)}}{\alpha_0^2}
  \del_{\tilde{E}}\tilde{n}_{t}(\tilde{E})|_{\tilde{E}=\tilde{\mu}(\tilde{t})}.
  \label{PopulationSlopeEq}
\eeq
The normalization condition (\ref{normalization}) then relates the posistion of the boundary to the slope of $\tilde{n}_{t}(\tilde{E})$ at the boundary:
\beq
  \del_{\tilde{E}}\tilde{n}_{t}(\tilde{E})|_{\tilde{E}=\tilde{\mu}(\tilde{t})}
  \approx -\alpha_0^2 e^{-\alpha_0 \tilde{\mu}(t)}.
  \label{SlopeCondition}
\eeq

\end{itemize}

\section{The propagating pulse}
\label{PropPulse}

\begin{figure}
\includegraphics[width=\columnwidth]{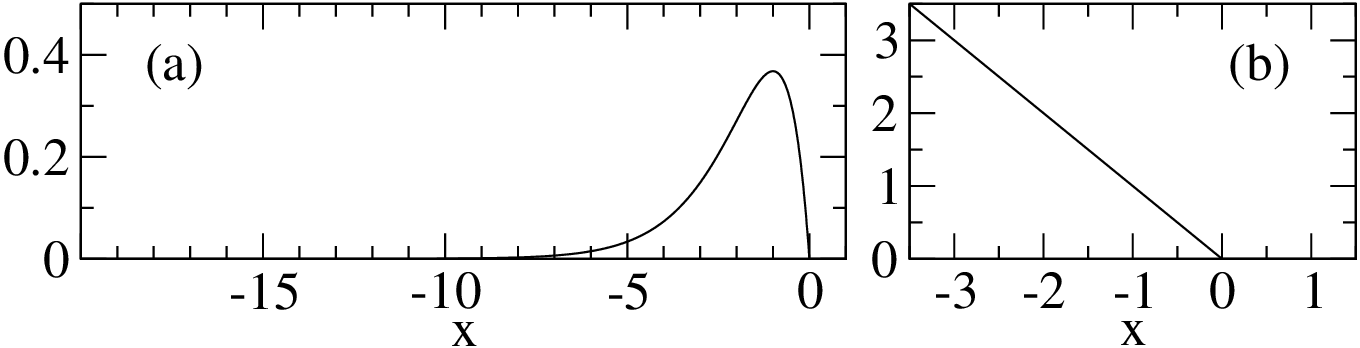}
\caption
{(a) The universal shape $f_0(x)$ of the propagating pulse. (b) The rescaled formulation $\tilde{f}_0(x) = f(x)e^{-x}$ for which the mean-field dynamics are purely diffusive.}
\label{univpulseshapefig}
\end{figure}

As argued above, $\tilde{n}_{t}$ will usually be very smooth on the characteristic energy scale $1/\alpha_0$, thus a linear expansion of $\tilde{n}_{t}$ around the boundary will give a good description of the population at time $t$. Applying Eq.~(\ref{SlopeCondition}), we find
\beq
  n_t(E) \approx \alpha_0 f_0(\alpha_0 [E-\mu(t)]),\label{pulseshape}
\eeq
where
\beq
  f_0(x) = \left\{\begin{array}{ll} -xe^x & x\leq 0\\ 0 & x>0\end{array}\right.
\eeq
is the universal pulse shape (Fig.~\ref{univpulseshapefig}). The dynamics of the system can now be described as the motion $\mu(t)$ of a pulse of almost constant shape. In the limit of large $t$ (and mean field), the pulse propagates at the constant speed $c_0$.

\subsection{Finite time}

A population that is located in a small energy range will initially propagate slower than the maximum speed $c_0$. As described above, the pulse shape is almost constant except for at very early times, thus the speed difference is equivalent to the motion of the boundary in Eq.~(\ref{diffusion}). This is easily modeled, to the lowest order correction: For a fixed boundary, Eq.~(\ref{diffusion}) has a straightforward solution, with a peak of height $\sim \frac{1}{t}$ a distance $\sim \sqrt{t}$ from the boundary, for which the slope at the boundary is $\sim \frac{1}{t^{3/2}}$. [Specifically,
$\tilde{n}_{t}(\tilde{E}) = \frac{1}{\gamma t} g(\frac{\tilde{E}}{\sqrt{\gamma t}})$, where $g(x)=-x e^{\frac{-x^2}{4}}$.]
The normalization condition given by Eq.~(\ref{SlopeCondition}) now gives the correct location of the boundary, $\tilde{\mu}(t) \sim \frac{3\ln(t)}{2\alpha_0}+\tilde{\mu}_0$, and the speed reduction is simply the time derivative of this. The logaritmic dependence on time indicates that the error caused by intially fixing the boundary is negligible for large $t$ ($\frac{\del}{\del t}\sqrt{\gamma t}\gg \frac{\del}{\del t}\frac{3\ln(t)}{2\alpha_0}$), but for small $t$ this error will be significant.

\begin{figure}
\includegraphics[width=\columnwidth]{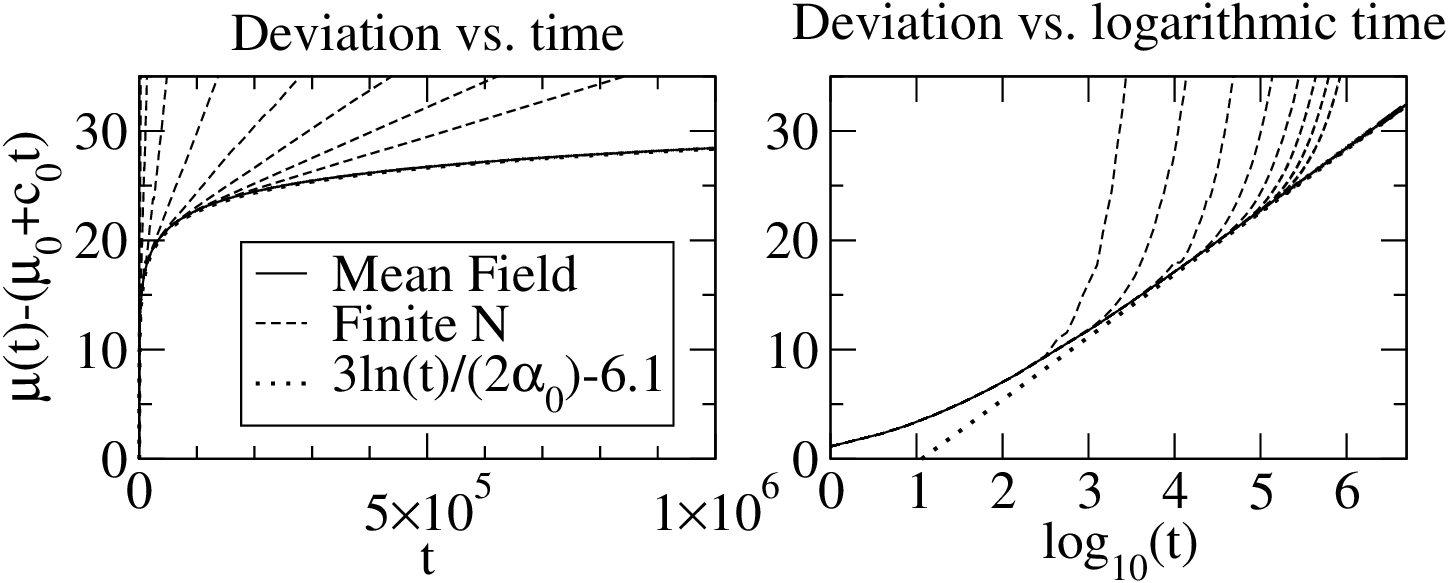}
\caption
{Finite time corrections to boundary position; mean field and finite population size simulations compared to theory prediction (plus arbitrary constant). Population sizes (left to right): $N=2^{20}$, $2^{50}$, $2^{100}$, $2^{200}$, ... , $2^{600}$.}
\label{FiniteTimeTrace}

\end{figure}
Agreement with simulations is excellent (Fig.~\ref{FiniteTimeTrace}); for finite populations, the speed settles down at $c_N$ once the pulse width $\sim \sqrt{\gamma t}$ reaches $\bar{\Delta}_N$ (see section~\ref{FinitePop} below).

\subsection{Discrete binding energies}

While deriving the diffusion Eq.~(\ref{diffusion}), we assumed that the DNA could have any binding energy. However, many models, as well as my simulations, only consider discrete values $i\epsilon_0$ of the binding energy, corresponding to $i$ mismatches from the WT sequence. Equation~(\ref{pulseshape}) clearly cannot be the correct pulse shape for the discrete case, as the relevant normalization condition $\sum_{i}n_t(i\epsilon_0)=1$ is satisfied only for discrete values of $\mu(t)$.

A more convenient approach for the discrete case is to consider the {\it cumulative} population distribution $h_t(E)=\int_{-\infty}^E n_t(E^\prime)dE^\prime$. From Eq.~(\ref{ODE}) it follows that $h_t(E)$ also obeys Eq.~(\ref{ODE}), but with a different boundary condition $\del_E n_t(E)|_{E=\mu(t)}=0$ and the simple normalization condition $h(\mu(t))=1$. We can perform the same transformations as before [$h_t(E)=\tilde{h}_t(E)e^{\alpha_0(E-c_0 t)}$ etc.] and replace the boundary condition with a ceiling:
\beq
  \del_{t}\tilde{h}_{t}(\tilde{E})
    \approx \gamma\del_{\tilde{E}}^2\tilde{h}_{t}(\tilde{E}),\;\;\;\;\;\;
  \tilde{h}_{t}(\tilde{E}) \leq e^{-\alpha_0 \tilde{E}} \;\;\;\forall \tilde{E},
  \label{cumulativediffusion}
\eeq
which no longer requires a separate equation for the chemical potential. Once we have the solution for $\tilde{h}_{t}(\tilde{E})$, and thus $h_t(E)$, we can find the correct population distribution given any discrete set $E_i$ of binding energies (we assume that the energies are ordered, $E_i<E_{i+1}$, for convenience): $n_t(E_i) = h_t(E_i)-h_t(E_{i-1})$.

The mutation rates $p(\epsilon)$ must apply equally to all the DNA in the population and must be zero unless $\epsilon$ is an integer times $\epsilon_0$. However, if the minimum change in binding energy (due to mutations) is $\epsilon_0$, then there is no way to enforce the boundary condition $\del_{\tilde{E}} \tilde{h}_{t} (\tilde{E})|_{\tilde{E}=\tilde{\mu}(t)} =\del_{\tilde{E}}e^{-\alpha_0 \tilde{E}}$, which follows from Eq.~(\ref{cumulativediffusion}), on any finer scale\footnote{The approximation I used between Eqs.~(\ref{randwalk}) and (\ref{diffusion_0}) gives a correction to the solution near the boundary even for the continuum case.}, and options 1 and 2 in Fig.~\ref{DiscreteBoundaryCondition} are a priori equally valid solutions. More precisely, these are the two extreme possibilities for average slope, and the real solution will be somewhere in between:

\begin{figure}
\includegraphics[width=\columnwidth]{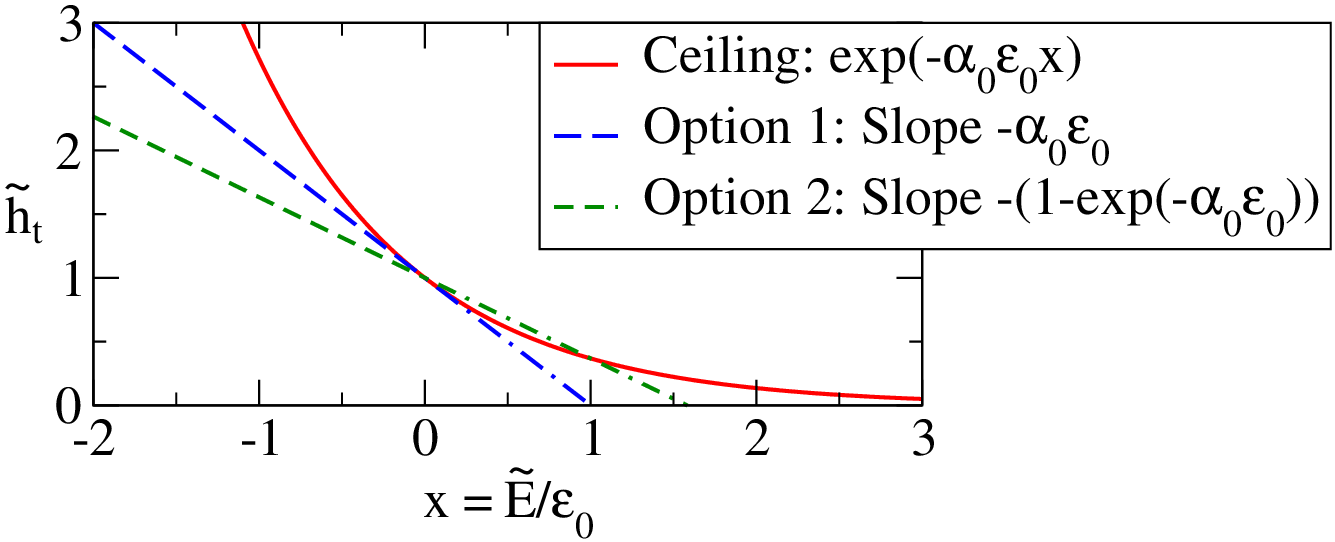}
\caption
{Alternate acceptable solutions to a continuous derivative boundary condition at $x=0$ for discrete dynamics. $\alpha_0\epsilon_0=1$.}
\label{DiscreteBoundaryCondition}
\end{figure}

\beq
  n_t(E) = h_t(E)-h_t(E-\epsilon_0) \approx f_{\alpha_0\epsilon_0}\left(\frac{E-\mu(t)}{\epsilon_0}\right),
\eeq
\beq
  f_{y}(x) = \left\{\begin{array}{ll}
  	(1+dx)e^{yx}-(1+d[x-1])e^{y(x-1)} & x\leq 0\\
  	1-(1+d[x-1])e^{yx} & 0<x\leq 1\\
  	0 & x>1
  	\end{array}
  	\right.
\eeq
where
\beq
  -y \leq d \leq -(1-e^{-y}).
\eeq
Note that $\mu(t)$ is here the binding energy at which $h_t(E)$ reaches 1, not the value at which $n_t(E)$ reaches 0 (they are identical for the continuous case, but for the discrete case, at zero temperature, the real chemical potential is not a continuous function of time).

\begin{figure}
\includegraphics[width=\columnwidth]{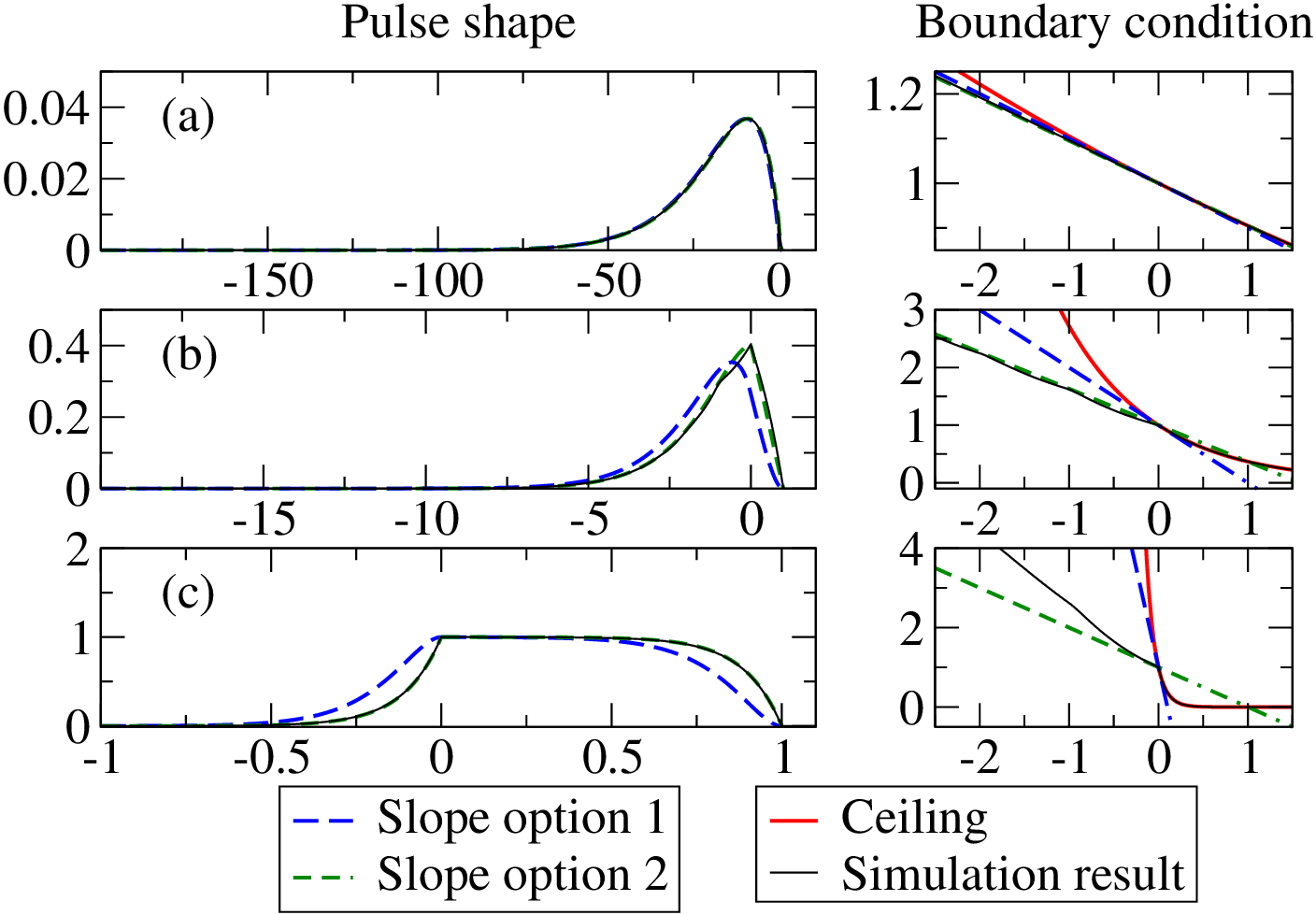}
\caption
{Numerical results and analytical estimates for discrete binding energies (see text). Left side: The pulse shape $n_t$ vs. $x=(E-\mu(t))/\epsilon_0$. Right side: The boundary condition for $\tilde{h}_t$ vs. x. (a)~$\alpha_0\epsilon_0 = 0.1$, $c_0\epsilon_0/\gamma=-0.25$. (b)~$\alpha_0\epsilon_0=1$, $c_0\epsilon_0/\gamma=-0.025$. (c)~$\alpha_0\epsilon_0=10$, $c_0\epsilon_0/\gamma\approx-2$.}
\label{pulseshapefig}
\end{figure}

For small $\alpha_0\epsilon_0$, the two extrema are close, and the pulse shape closely matches that of the continuum solution [Fig.~\ref{pulseshapefig}(a)]. For large $\alpha_0\epsilon_0$, the upper boundary $d=-(1-e^{-\alpha_0 \epsilon_0})$ gives the better approximation, although any acceptable value of $d$ gives a qualitatively correct solution [Fig.~\ref{pulseshapefig}(c)].

This approximation involves two dimensionless parameters: the primary parameter $\alpha_0\epsilon_0$ comes from specifying the energy scale $\epsilon_0$, but there is also a secondary parameter $c_0\epsilon_0/\gamma$ which comes from specifying a unique ``at rest" coordinate system. The slope $d=-(1-e^{-\alpha_0\epsilon_0})$ is the limit for $c_0\epsilon_0/\gamma\rightarrow 0$.

A large value of $\alpha_0\epsilon_0$ corresponds to low mutation rate and strong selection. Given the constant propagation speed from Eq.~(\ref{LowMRPropSpeed}), we can also consider the number of DNA molecules with a given energy $E_i$ as a function of time: $N_i(t) = N f_{\alpha_0\epsilon_0}(x_{i,0}-c_0t/\epsilon_0)$. For large $\alpha_0\epsilon_0$ (and $c_0$ small and negative), the general behavior of this function agrees with the argument in~\cite{KloTan} [Fig.~\ref{pulseshapefig}(c)]: $N_i$ grows exponentially until $N_i\approx N$, {\it i.e.}, almost the entire population has been replaced, and then remains almost constant until the population is replaced again.

\section{Finite population size}
\label{FinitePop}

As discussed in the introduction, the dependence of the propagation speed on the population size is of significant interest. In reference~\cite{PenGerHwaLev}, Peng {\it et al.} estimate the effects of the population size by using a cutoff procedure~\cite{BruDer} that ignores amplification for the part of the distribution where $n(E)<\frac{1}{N}$.\footnote{A better version of this cutoff uses $h(E)<\frac{1}{N}$, which does not depend on the energy scale/resolution.} While this is a reasonable approach and gives a result that matches simulations fairly well, it is not clear whether this is the correct cutoff~\cite{BruDer}.

\begin{figure}
\includegraphics[width=\columnwidth]{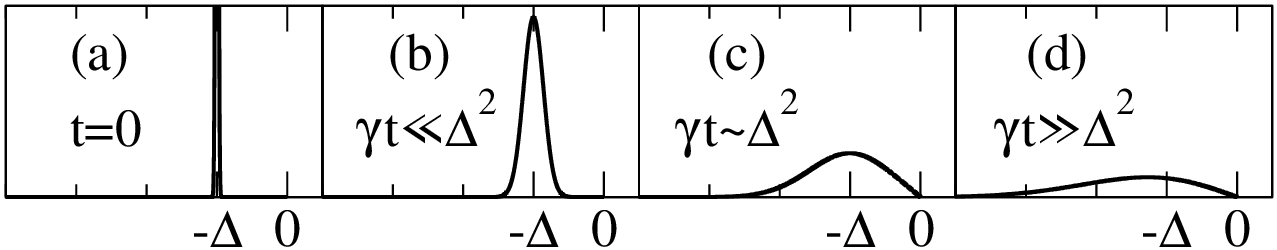}
\caption
{An initial perturbation given by a delta function (a) will spread as a Gaussian (b) until it encounters the boundary. As some time $\sim \Delta^2$ the slope at the boundary will be maximal $\sim 1/\Delta^2$ (c), after which it decays as $\sim t^{-3/2}$ for $t\rightarrow\infty$ (d).}
\label{EvolutionOfPerturbation}
\end{figure}

\begin{figure}
\includegraphics[width=\columnwidth]{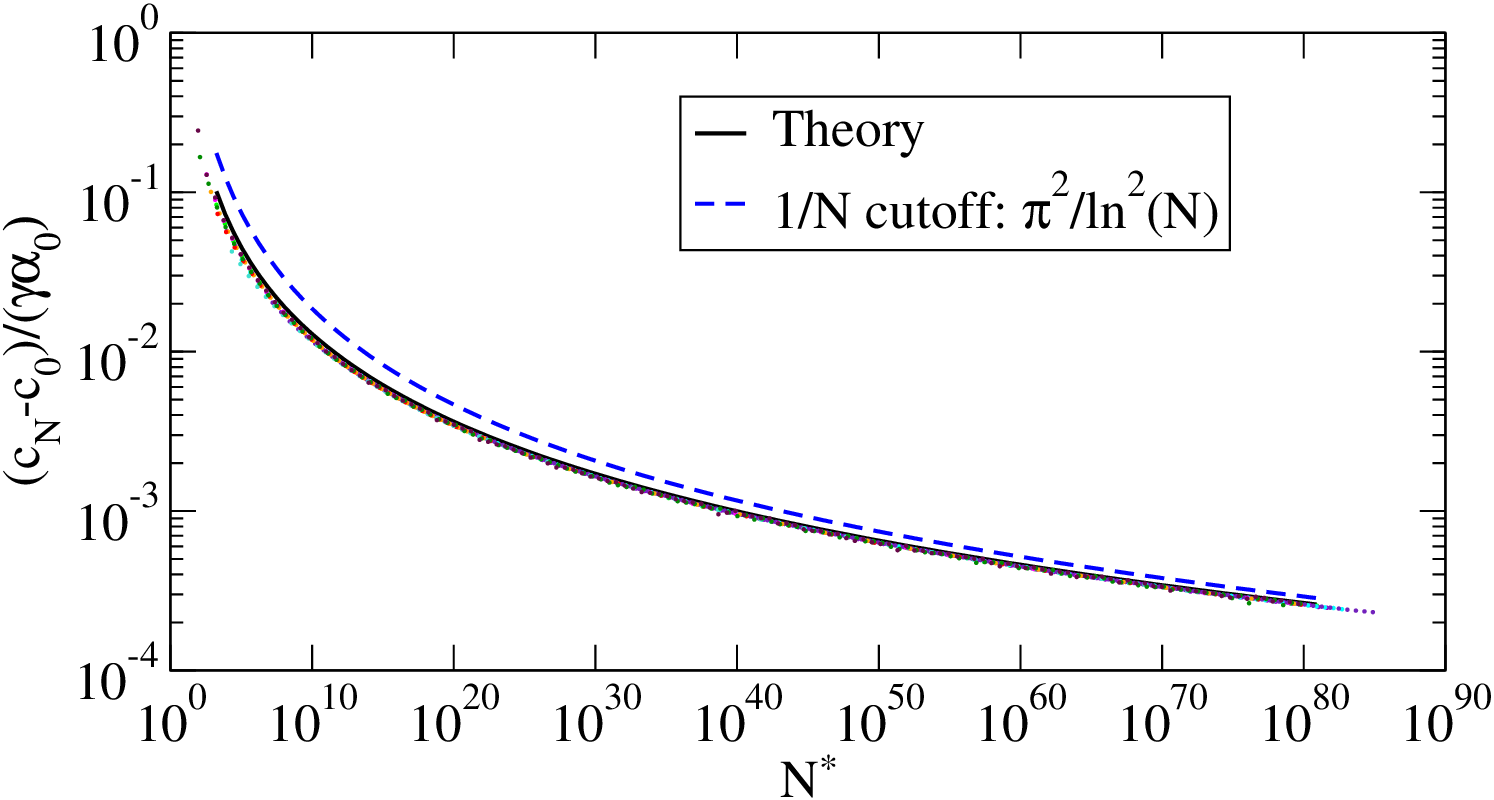}
\caption
{Deviation from mean-field propagation speed as a function of effective population size $N^*$ (see text); theory results and scaling collapse of simulations (small circles).}
\label{PRLSpeedChange}
\end{figure}

A better approach is to base the cutoff procedure explicitly on the key criterion: when will mean-field theory fail? Consider what happens to a single DNA molecule with a binding energy $\Delta$ less than the selection threshold $\mu_0$, according to mean-field theory. The molecule corresponds to an initial perturbation $\delta n_0(E)=\frac{1}{N}\delta(E-[\mu_0-\Delta])$, where $\delta(\cdot)$ is Dirac's delta function, and the transformed perturbation $\delta\tilde{n}_0(E)=\frac{1}{N}e^{\alpha_0 (\Delta-\tilde{\mu}_0)}\delta(E-[\tilde{\mu}_0-\Delta])$ will evolve according to Eq.~(\ref{diffusion}) with a fixed boundary $\tilde{\mu}(t)=\tilde{\mu}_0$. Simple scaling requires that the slope at the boundary, $\frac{\del \tilde{n}_t(\tilde{E})}{\del \tilde{E}}|_{\tilde{E}=\tilde{\mu}_0}$, reaches a maximum value
$\sim \frac{1}{N}e^{\alpha_0 (\Delta-\tilde{\mu}_0)}/\Delta^2$ after a time $t_{\Delta}^{\rm max}\sim \Delta^2/\gamma$, as shown in Fig~\ref{EvolutionOfPerturbation}.
According to Eq.~(\ref{PopulationSlopeEq}), this corresponds to a population size perturbation of
$\delta N_{\Delta}^{\rm max} \sim e^{\alpha_0 \Delta}/(\alpha_0 \Delta)^2$
descendants (for $\alpha_0\Delta\gg 1$). For $\Delta$ large enough, $\delta N_{\Delta}^{\rm max}>N$, {\it i.e.}, the descendants of a single molecule will eventually replace the whole population, and the selection threshold must ``jump" to keep the population size constant. This is a clear violation of mean field, and the condition $\delta N_{\Delta_N}^{\rm max}=N$ gives a lower boundary $\mu_2(t) = \mu(t)-\Delta_N$ beyond which mean field fails. The best estimate of the propagation speed that we can find from mean field is then the propagation speed $c_{{b}(\bar{\Delta}_N)}$ of a pulse of length
$\bar{\Delta}_N = \Delta_N-t_{\Delta_N}^{\rm max}(c_{{b}(\bar{\Delta}_N)}-c_0)$, where ${b}(\bar{\Delta}_N) = \pi/\bar{\Delta}_N$; this includes a correction for the fact that the boundary in eq.~(\ref{diffusion}) would actually move:\footnote{Numerically, $t_{\Delta}^{\rm max}\approx \Delta^2/(6\gamma)$.}
\beq
  c_N \approx c_0 + \frac{\gamma\alpha_0\pi^2}{(\ln[N\ln^2(N)]-\pi^2/6)^2}.
  \label{FinitePopSpeedEq}
\eeq
As shown in Fig.~\ref{PRLSpeedChange} (``Theory"), this estimate is very accurate. It slightly overestimates the correction, since it ignores the contribution from the mean-field-violating events, but it is far better than the estimate in~\cite{PenGerHwaLev} (``1/N cutoff").

\begin{figure}
\includegraphics[width=3in]{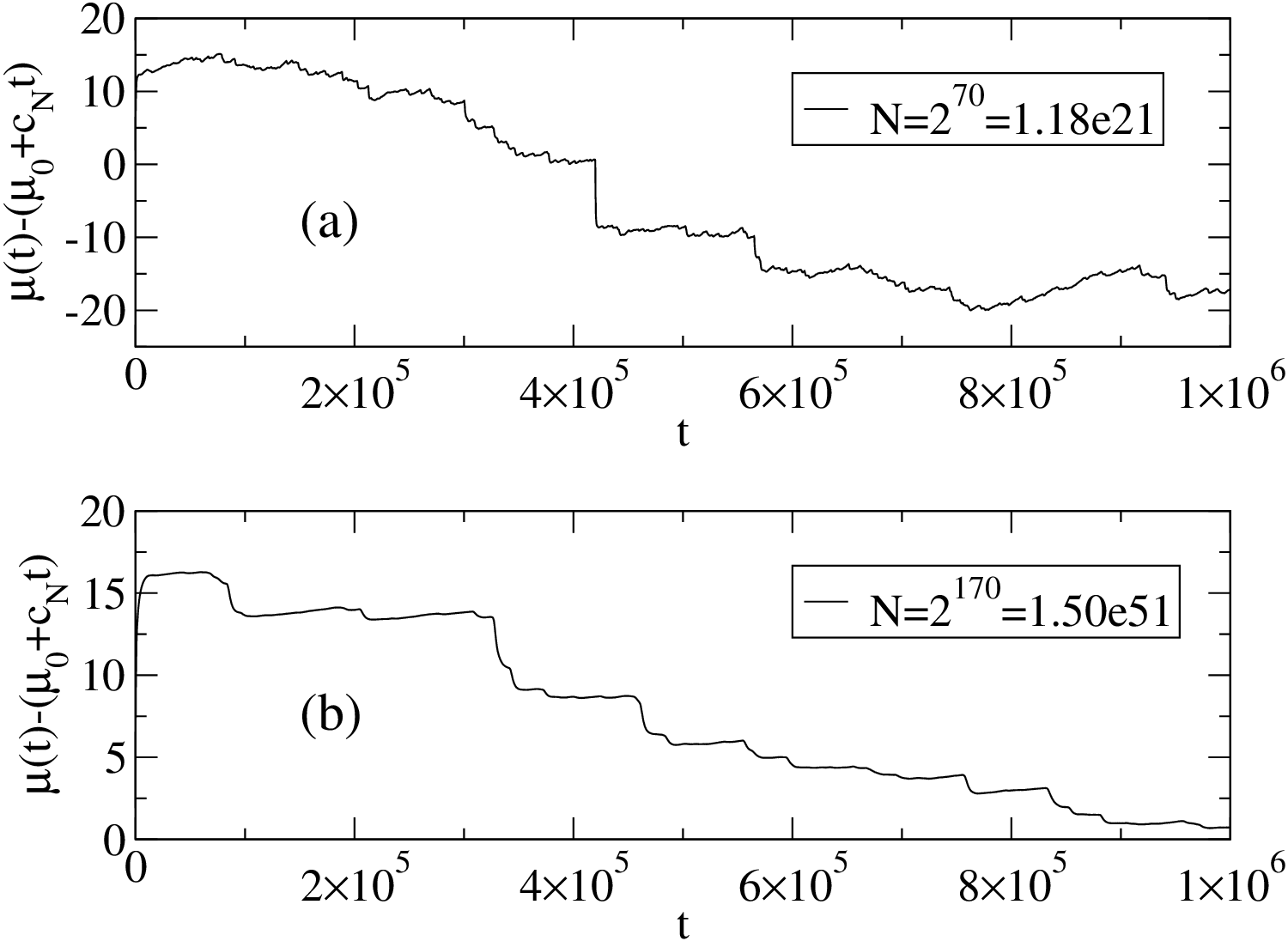}
\caption
{Motion of boundary $\mu(t)$ relative to estimate [Eq.~(\ref{FinitePopSpeedEq})] for simulations with different population sizes.}
\label{PRLsimrandomevents}
\end{figure}

Fig.~\ref{PRLsimrandomevents} shows the difference between the actual value of $\mu(t)$ and our estimate $\mu_0+c_N t$ as a function of time for two simulations with different populations sizes. These plots support our discussion above---the difference remains almost constant for long periods, but then it suddenly makes a large jump, {\it i.e.}, there is an event that violates the mean-field assumption. Note that all jumps of size less than $\sim \frac{1}{\alpha_0}$ are included in the estimate $c_N$. The larger population size has a much smoother behavior than the smaller one, which is expected since the time required to propagate a random event scales as $t_{max}\sim \Delta_N^2\sim\ln(N)^2$.

\subsection{Effective population size}

The argument above is based on the continuous time formulation of the evolution process. There are two types of corrections to this formulation: The selection is only done at discrete intervals, which applies also to mean field, and the mutation and replication may be done in various different ways, which changes the statistical properties of the random noise.

\begin{figure}
\includegraphics[width=2in]{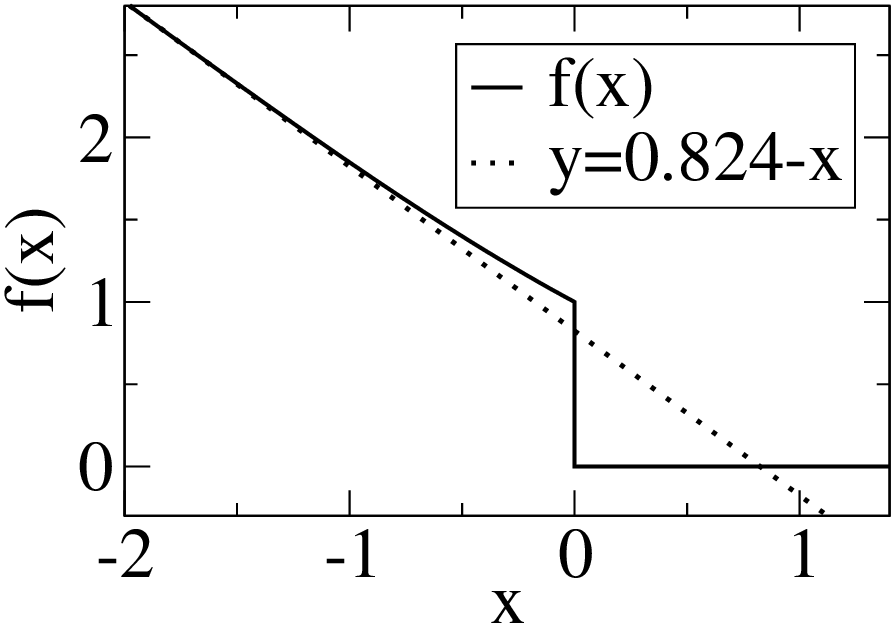}
\caption
{Effective boundary for discrete selection (see text).}
\label{DiscreteSelectionFig}
\end{figure}

The selection process only affects the tail of the propagating pulse. If we solve the diffusion equation $\frac{\del f(x,t)}{\del t} = \frac{\del^2 f(x,t)}{\del x^2}$ numerically with the boundary condition $f(x,t)= -x$ for $x\ll 0$ and set $f(x,t)$ to zero for $x>0$ at integer values of $t$, the function at integer values of $t$ will be as shown in Fig.~\ref{DiscreteSelectionFig}. The effective boundary position is at $x\approx 0.824$, and $\frac{f(0)}{0.824}\approx 1.21$. This yields an effective population size (the population size corresponding to the dotted line in Fig.~\ref{DiscreteSelectionFig}) $N^\prime \approx N\frac{e^{\alpha_0\delta}}{1+1.21\alpha_0\delta}$, where $\delta=0.824\sqrt{\gamma}$---this is accurate for $N$ large enough that $\tilde{n}(\tilde{E})$ has almost constant slope at least a distance $\delta$ from the boundary, {\it i.e.}, for $N\gg \frac{N^\prime}{N}$.

If replication and mutation are a single process, for instance error-prone PCR, then there is no further correction---this corresponds to the continuum formulation. If, however, the mutations are introduced separately, after replication, then the population essentially evolves according to mean field between selections, and the second boundary behaves the same way as the first boundary, {\it i.e.}, the effective cutoff is a distance $\delta$ from the nominal value. The actual population by the second boundary is negligible, thus the effective population is simply $N^{*} = N^\prime e^{\alpha_0\delta}$.

The high quality of the scaling collapse in Fig.~\ref{PRLSpeedChange} shows that the second boundary $\mu_2(t)$ very accurately captures the effects of a finite population on the propagation speed: Figure~\ref{PRLSpeedChange} includes simulations for which the $N^*/N^\prime$ corrections, which are entirely due to the second boundary, vary from close to 1 to about 1600, and the collapse is accurate to better than a factor of 2 in the population size. Note that $N^*/N^\prime$ may be larger than the amplification factor $K$, thus the effective population size can be larger than the largest population that occurs during the evolution, $KN$.

\section{Finite length}
\label{FiniteLength}

The strongest assumption used to derive Eq.~(\ref{diffusion}) was that the mutation rate $p(\epsilon)$ is independent of the initial binding energy, or, more generally, the genotype of the molecule that is mutated---I have yet to prove that this formulation can capture the behavior of finite systems. While it is difficult to prove this in general, since the above assumption was precisely what allowed the use of an otherwise fully general mutation rate $p(\epsilon)$, we can again consider the cases of high/low mutation rate and weak/strong selection.

\subsection{High mutation rate/weak selection}
Consider the discrete, finite model described in section~\ref{Simulations}. As mentioned in section~\ref{OldResults}; for high mutation rate and weak selection, it is convenient to describe the mutation rate of an infinite system in terms of its diffusion coefficient $D$ and drift speed $v$ [Eq.~(\ref{D_and_v_mutation_rate})]. Using the same approach in a finite system, $D$ and $v$ will depend on the number of mismatches $r$~\cite{PenGerHwaLev}, thus the constant speed solutions we have found will not be valid. However, as long as the parameters vary slowly on the scale of the characteristic length $\frac{1}{\alpha_0}$ of the infinite model, we expect these solution to be good local approximations.

\subsubsection{Exponential approach to equilibrium}
\label{equilibriumapproach}

In~\cite{PenGerHwaLev}, Peng {\it et al.} argued that for small $k$, an initial population that is compact in mismatch space and far from equilibrium will form a moving pulse that exponentially approaches the equilibrium position; $r_0(t)-r_0^{EQ}\sim e^{-\frac{A}{A-1}\nu_0 t}$, where $r_0(t)=\mu(t)/\epsilon_0$ is the position of the boundary in mismatch space. If we numerically solve Eqs.~(\ref{Deq}--\ref{keq}) for $\alpha_0$, $c_0$ and $\gamma$ as a function of $r$, while keeping $k$ constant, we indeed find that for a sizable range of $k$, $c_0$ depends almost exactly linearly on $r$, thus
\beq
  \frac{\del r_0(t)}{\del t} \approx c_0(r_0(t)) \approx \frac{\del c_0(r)}{\del r} [r_0(t)-r_0^{EQ}],
\eeq
which leads to $r_0(t)-r_0^{EQ}\sim e^{-\frac{\del c_0(r)}{\del r}t}$. The exponential constant $-\frac{\del c_0(r)}{\del r}$ agrees well with simulations (Fig.~\ref{rescaleplot}) and will in general differ from the estimate $-\frac{A\nu_0}{A-1}$ found in~\cite{PenGerHwaLev}, which is the limit for $k\rightarrow 0$.

Surprisingly, the linear dependence of $c_0(r)$ on $r$ is most accurate not for very small $k$; rather, there is a finite value of $k$ at which $c_0(r)$ and $\gamma(r)$ are perfectly linear, while $\alpha_0$ is constant and given by the solution to\footnote{Solve the $r$-derivatives of Eqs.~(\ref{Deq}--\ref{keq})
using the conditions $\frac{\del D}{\del r}=-\frac{\nu_0}{2}\frac{A-2}{A-1}$, $\frac{\del v}{\del r} = -\nu_0\frac{A}{A-1}$ and $\frac{\del k}{\del r}=0$.}
\beq
  1-(1+\alpha_0)e^{-\alpha_0} = \frac{2}{A}(1+\alpha_0\sinh{\alpha_0}-\cosh{\alpha_0}).
\eeq
This would give a perfectly exponential approach to equilibrium of a perfectly shape-preserved pulse at this single value of $k$ (if we ignore corrections from finite time, discrete mismatch numbers, etc.)---the variation of $\gamma$ does not affect either shape or movement once the pulse has reached its constant shape. This is also the value of $k$ that gives the smallest $\frac{\del c_0(r)}{\del r}$, {\it i.e.}, the slowest convergence. Simulation results verify that the exponential approach is near perfect for this $k$, and better than for either weaker or stronger selection (not shown).

\subsubsection{Equilibrium position}
The simplest estimate we can make for the equilibrium value $r_0^{EQ}$ is the position at which the values of $D$ and $v$ are such that the propagation speed $c_0$ would be 0 [from Eqs.~(\ref{Deq}--\ref{keq})]:
\beqar
  k & = & 2D(r)-\sqrt{4D(r)^2-v(r)^2}. \label{EqPosLowerBound}
\eeqar
However, the entire population has lower mismatch number than $r_0$, which means that the solution to Eq.~(\ref{EqPosLowerBound}) is a lower bound for $r_0^{EQ}$.

\begin{figure}
\includegraphics[width=\columnwidth]{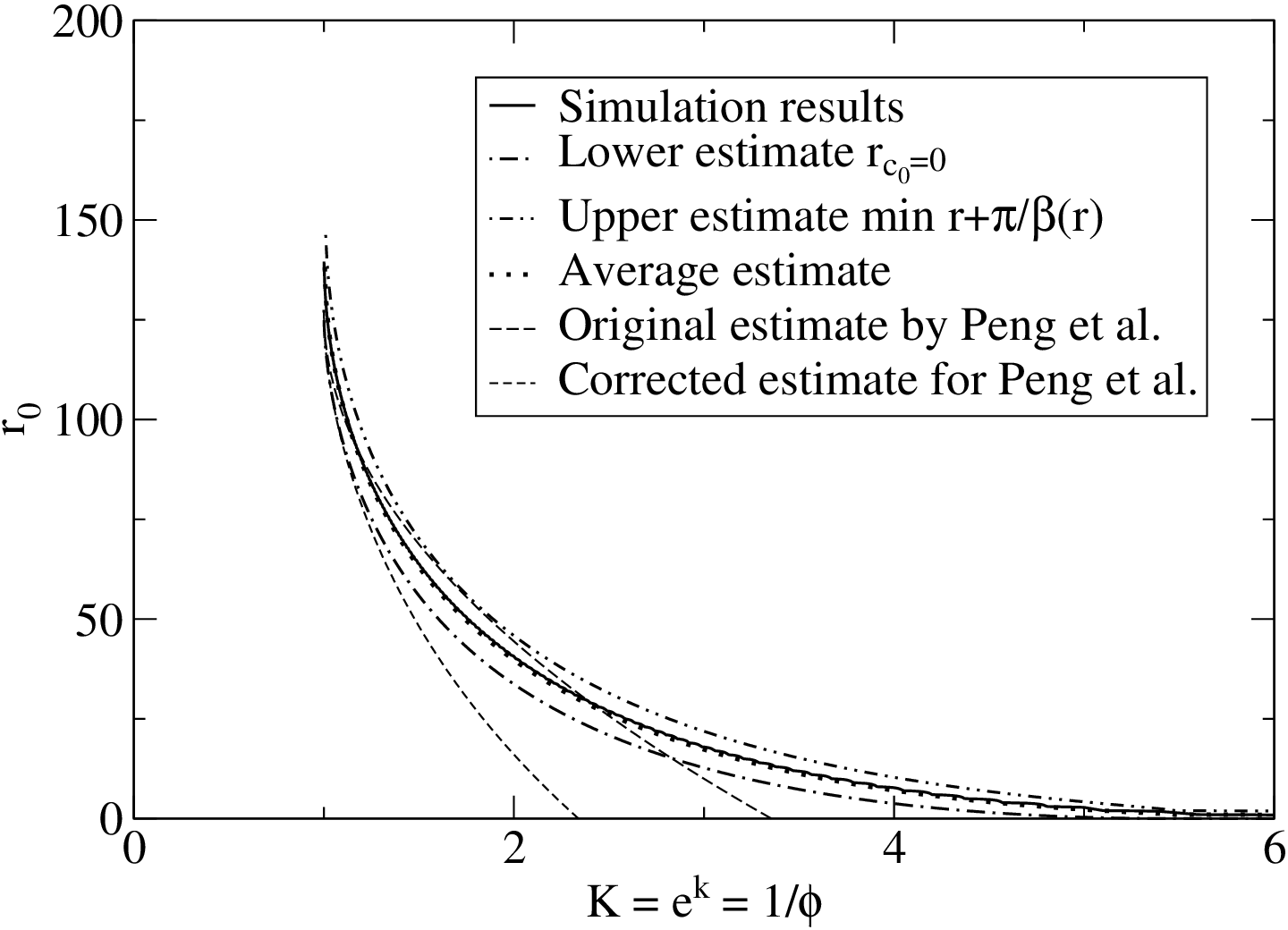}
\caption
{Equilibrium position for the boundary as a function of amplification K; estimates and simulation results, as discussed in the text. $L=170$, $\nu=0.01$.}
\label{PRLEqFig}
\end{figure}

Imposing the equilibrium condition $c_{{b}}=0$ on a {\it bounded} solution yields
\beqar
  k & = & 2D(r)-\sqrt{4D(r)^2-v(r)^2}\cos({b}).
\eeqar
By solving the above equation for ${b}$, we find a population that lies between $r$ and $r+\frac{\pi}{{b}(r)}$, {\it i.e.}, the entire population has {\it higher} mismatch number than the position at which the equilibrium parameters occur. Its upper bound $r+\frac{\pi}{{b}(r)}$ is thus an upper bound for $r_0^{EQ}$, and to find the best upper bound we minimize over $r$. Our estimate for the equilibrium value of $r_0$ is then the average of the lower and upper bound:
\beq
  r_0^{EQ} \approx \frac{r_{c_0=0}}{2}+\frac{1}{2}\min_{r>r_{c_0=0}}r+\frac{\pi}{{b}_{c_{b}=0}(r)}
\eeq
As shown in Fig.~\ref{PRLEqFig}, this estimate is very accurate for most values of $\phi^{-1}=K=e^k$; for $K>1.02$, the discrepancy is always less than 1.\footnote{For $K$ very close to 1, $\alpha_0$ is small, thus the assumption that $D(r)$ and $v(r)$ vary slowly on the scale of $\frac{1}{\alpha_0}$ does not hold.} The lower bound given by Eq.~(\ref{EqPosLowerBound}) is equal to the ``second region formula" in~\cite{CohKes}. In Fig.~\ref{PRLEqFig}, I also include the estimate found in~\cite{PenGerHwaLev} for comparison, as well as a ``corrected" version in which $\frac{\phi^{-1}-1}{\tau}$ has been replaced by $k=\ln{\frac{1}{\phi}}$ (as in section~\ref{OldResults}); this second version is very close to the lower bound for small $k$.

\subsection{Low mutation rate/strong selection}

Reference~\cite{KloTan} describes simulations of {\it in vitro} molecular evolution based on a realistic model of the binding energy of a DNA sequence and the {\it Mnt} protein and found that the evolution rate, in terms of number of iterations required to reduce the average number of mismatches by one, was very close to constant in the mean-field limit. According to Eq.~(\ref{LowMRPropSpeed}), the propagation speed should depend logarithmically on the beneficial mutation rate $p_-$, in which the number of remaining mismatches $m$ is a factor. However, this equation describes the rate of evolution of binding energy, not of the number of mismatches, and it only holds when all beneficial mutations confer the same decrease in binding energy.

It is easy to show that decreasing the rate/benefit of beneficial mutations and increasing the rate/penalty of deleterious mutations will always decrease the energy evolution rate: $-c_0(p_2)\leq -c_0(p_1)$ whenever $\int_{-\infty}^E \epsilon p_2(\epsilon)d\epsilon\geq\int_{-\infty}^E \epsilon p_1(\epsilon)d\epsilon\;\forall E$. This does not hold for the mismatch evolution rate
\beq
  \frac{\del r(t)}{\del t} = \sum_{i\in M_-} e^{-\alpha_0\epsilon_i(S)}\frac{\nu_0}{3} - \sum_{i\in M_+} e^{-\alpha_0\epsilon_i(S)}\frac{\nu_0}{3},
\eeq
where $M_\pm$ are the sets of mutations that would correct a mismatch ($-$) / introduce a mismatch ($+$), and $\frac{\nu_0}{3}$ ($=\frac{r}{6}$ in~\cite{KloTan}) is the rate of each specific mutation. If there is a single mismatch that is associated with a much larger energy penalty than any other, then mutants that have corrected this mismatch will outcompete all other mutants, even ones that have corrected multiple other mismatches. This may lead to a lower initial rate of mismatch correction than if there was no such extreme mismatch, or if it could not be corrected.

For realistic models such as the one in~\cite{KloTan}, the various mismatches are associated with a wide range of energy penalties. If the mutation rates are similar for the different mismatches, then the mismatch evolution rate may indeed be very close to constant during most of the evolution. For the intial population used in~\cite{KloTan}, which has 6 mismatches, the effective number of mismatches (the number of mismatches that would give the observed mismatch evolution rate if they all had the same energy penalty) is less than 2.5 for a mutation rate of $10^{-7}$ (the lower the mutation rate, the lower the effective number of mismatches).

\section{Finite temperature}
\label{FiniteTemp}

So far we have used perfect selection, {\it i.e.}, we always keep the best $N$ protein binders. This is a good approximation if $\beta\sqrt{\gamma}\gg 1$, or, for discrete models, $\beta\epsilon_0\gg 1$. The more realistic situation of finite temperature is far more difficult to model, as we can no longer take the continuum limit in the formal manner described in appendix~\ref{continuumapprox}. There are, however, two relatively simple limits.

For sufficiently large populations, or in mean field, the population distribution will extend to energies much lower than the chemical potential, where the selection factor $\frac{1}{1+e^{\beta[E(S)-\mu]}}$ is practically equal to one. In this region, the behavior is still governed by Eq.~(\ref{diffusion}). We can consider the region around $\mu(t)$ to impose a non-trivial, but fixed, boundary condition, and the evolution rate will still be $c_0$ for mean field ($c_{N^*}$ for finite populations, where the first effective population size correction $N^\prime/N$ depends on the temperature, but not on $N$).

On the other hand, for finite populations and very strong selection, all of the population will have binding energy much {\it higher} than $\mu(t)$. We can then approximate
\beq
  P(E) = \frac{1}{1+e^{\beta(E-\mu)}} \approx e^{\beta(\mu(t)-E)}.
  \label{LargeKSelectionApprox}
\eeq
Here, the relative fitness of two molecules is constant; $P(E_1)/P(E_2)=e^{\beta(E_2-E_1)}$; thus, this limit is equivalent to the traditional fixed-fitness models of evolution! The exponential dependence on $E$ again allows a formal continuum limit of the mean-field theory (see appendix~\ref{FiniteTempContinuum}). While this mean-field theory is not a good approximation for {\it any} finite population size~\cite{RidLevKes}, one can apply the same criterion as in section~\ref{FinitePop}, $\delta N^{\rm max}=N$, to find bounded-from-below solutions that capture the behavior of finite populations\footnote{The population size perturbation must here be calculated in the presence of the boundary.}. This is an alternative approach to the statistical criterion used in~\cite{RouWakCof}.

We can easily find the propagation speed of a single molecule (for the discrete evolution process): The selection probability is exponential in the energy, and this simply alters the effective mutation rate (mutating with rate $r$ and then selecting the mutants with probability $p$ times the probability of selecting the original is equivalent to mutation rate $rp$), giving the propagation speed
\beq
  c_1 = \int_{-\infty}^{\infty} \epsilon e^{-\beta\epsilon} p(\epsilon)d\epsilon.
  \label{SingleParticleSpeed}
\eeq
The propagation speed for a single molecule has the same form as the propagation speed $c_0$ in the limit of large populations, when the approximation in Eq.~(\ref{LargeKSelectionApprox}) no longer holds. This immediately shows that for $\beta>\alpha_0$, the approximation makes no sense.

\section{Discussion}

In the preceding sections, we have seen how the assumption of a translationally invariant mutation rate leads to a simple set of equations describing the mean-field theory of competitive evolution, and that this powerful formulation can give very accurate predictions in many different situations. The effective dynamics of the evolution process are to a good approximation captured by the rescaled mutation rate $\tilde{p}(\epsilon)=e^{-\alpha_0\epsilon}p(\epsilon)$. Far from equilibrium, or for small populations, this is the rate at which mutations are fixed in the population, while for a sufficiently large population close to equilibrium, this gives the relative occurrences of different bases: For an additive energy model, the equilibrium $p_i(b_2)/p_i(b_1)$ between two bases $b_1,b_2$ at position $i$ that have an energy contribution difference of $\delta E_i=E_i(b_2)-E_i(b_1)$ will be shifted by a factor $e^{-2\alpha_0 \delta E_i}$ relative to the equilibrium $p_i^0(b_2)/p_i^0(b_1)$ for pure mutation, without any selection.

One significant result was the accurate prediction of the limits of mean-field theory for a finite population and the way in which it fails. Namely, when a molecule by chance gets sufficiently far ahead of the general population, the population is quickly replaced by the descendants of that single molecule, evidenced by a jump in the selection threshold. This result has important consequences for the maximum genetic variation that can build up during competitive evolution: Whenever such a jump occurs, essentially all prior genetic variation is wiped out. For instance, for an organism that reproduces mostly by asexual growth\footnote{Where the asexual growth can be described by competitive evolution.} but with periodic sexual reproduction stages, this sets a lower limit for the frequency of sexual reproduction necessary to maintain genetic diversity.

Interestingly, the parameter values of the experiment performed by Dubertret {\it et al.}~\cite{DubLiuOuyLib} place it in the high-temperature regime described in section~\ref{FiniteTemp}, which has the same dynamics as the traditional fixed-fitness models of evolution: Even though the population is fairly large ($\approx 10^8$), the very large amplification ($2^{25}$) means that during selection, even the strongest protein binders may be removed. Also, the system used is only slightly larger than the propagating pulse, thus finite size effects will likely be severe (the pulse length is far more important in the high-temperature regime than in the $T=0$ limit).

While the global translational invariance of the model presented here is a very strong assumption, most of the results will be locally accurate for any system for which the mutation rates $p(\epsilon)$ change slowly on the characteristic length scale $1/\alpha_0$. As shown in section~\ref{FiniteLength}, one can fairly easily estimate corrections when the energy dependence of the mutation rate has a known, simple form, yielding even more accurate results.

In summary, I have presented a simple model that accurately captures the key qualitative and quantitative features of asexual competitive evolution on a smooth landscape. This allows prediction of the performance of competitive evolution for a wide range of parameters for any given, reasonably smooth energy landscape. Additionally, this forms a baseline against which the performance of more complex evolutionary procedures, {\it e.g.} using recombination~\cite{PenLevHwaKes}, can be compared.

\acknowledgments
The author would like to thank Rahul Kulkarni for helpful comments.

\appendix
\section{Continuous time approximation}

\label{continuumapprox}

As before, let $n_t(E)$ be the normalized population density at time $t$. Formally, an evolution cycle can be represented by the operators $\R=K$ (reproduction), $\M$ (mutation) and $\Se$ (selection):
\beq
  n_{t+1} = \Se \M \R n_t.
  \label{DiscreteIteration}
\eeq

\begin{figure}
\includegraphics[width=\columnwidth]{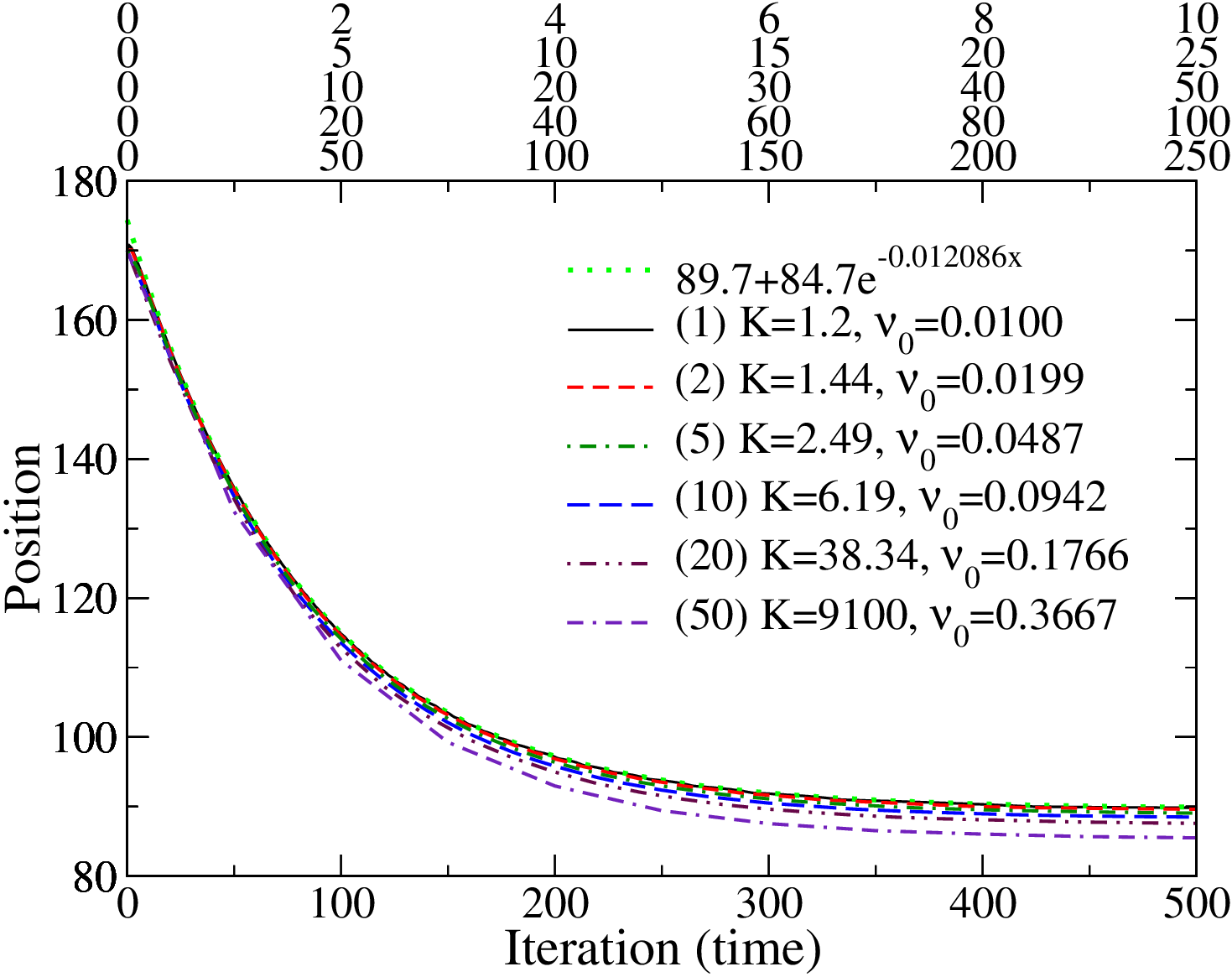}
\caption
{Simulations of different models in the same scaling class, plotted on appropriately scaled time axes. The scaling factor $l$ is in parantheses. For small $l$, the approach to equilibrium is very well fit by an exponential with the constant $-\frac{\del c_0(r)}{\del r}$, as expected (section~\ref{equilibriumapproach}).}
\label{rescaleplot}
\end{figure}

Both $\M$ and $\R$ can be written as the exponentials of other operators\footnote{Even in situations where $\M$ can not be written exactly as an exponential, this is a good approximation as long as the mutation rate per base per iteration is small.} [$\R=e^{\re}$, where $\re=k=\ln(K)$, and $\M = e^{\m}$], while $\Se=\Se^2$. This allows us to generalize Eq.~(\ref{DiscreteIteration}) to the approximate equation
\beq
  n_{t+\Delta t} = \Se e^{\Delta t \m} e^{\Delta t \re} n_t,
\eeq
which in the limit $\Delta t\rightarrow 0$ becomes
\beq
  \del_t n_t = (\m + \re) n_t\;,\;\;\;\;\;\;n_t[\mu(t)]=0
  \label{continuummodel}
\eeq
---since selection is done via a step function, the continuum limit is the boundary condition $n_t[\mu(t)]=0$, where $\mu(t)$ is dynamically chosen such that $n_t$ remains normalized. $\m$ and $\re$ commute, thus the only sources of error in this approximation are the commutators with the selection operator/boundary condition, and these are localized around the bondary. By rescaling time we can identify classes of models that are equivalent within this approximation:
\beqar
   t_l & = & t/l \\
   \R_l & = & e^{l \re} = K^l \\
   \M_l & = & e^{l \m} \\
   \Se_l & = & \Se.
\eeqar
The continuum model [Eq.~(\ref{continuummodel})] is a scaling limit of this class, and all models within the class that are sufficiently close to this limit ({\it i.e.}, $\R$ and $\M$ are close to 1) should be well approximated by it. Figure~\ref{rescaleplot} shows simulations of various models from the same scaling class; $\M$ is parametrized by $\nu_0$, which is rescaled according to $\nu_{0,l}=\frac{A-1}{A}[1-(1-\frac{A}{A-1}\nu_0)^l]$.

The continuum approximation fails if rescaling by a factor of 2 is no longer accurate, {\it i.e.}, if $\Se \M_{2\Delta t} \R_{2\Delta t} n_t$ is significantly different from $\Se \M_{\Delta t} \R_{\Delta t} \Se \M_{\Delta t} \R_{\Delta t} n_t$. The difference is that molecules that in one case are removed after $\Delta t$ in the other case will survive until time $2\Delta t$, but will then face a stronger selection. These molecules can only make a difference if they reach the leading (low $E$) edge of the population. As they start out at the tail of the population at time $\Delta t$, there are only two ways in which they can reasonably reach the leading edge of the population: Either the propagating pulse is very short, which happens only in finite populations, or the mutations are strong enough to almost equilibrate the distribution, {\it i.e.}, $\nu_0$ is large ($L$ is finite). In Fig.~\ref{rescaleplot}, the simulations deviate significantly from the continuum limit only for $\nu_0$ on the order of 1.

\subsection{Finite temperature}
\label{FiniteTempContinuum}

As noted above, the discrete evolution process can be represented using operators:
\beq
  n_{t+1} = \Se \M \R n_t = \Se e^{\m} e^{\re} n_t.
\eeq
For large temperatures, very strong selection and finite populations\footnote{Indeed: this is a mean-field theory that is only applicable to finite populations!}, Eq.~(\ref{LargeKSelectionApprox}) allows us to write the selection operator as $\Se=\N e^{-\beta\E}$, where the normalization operator $\N$ simply rescales a function; $\N f = \frac{f}{\int f(E) dE}$, and $\E$ is the binding energy operator; $\E f(E) = E f(E)$. The differential mutation operator $\m$ can be explicitly written as
\beq
  \m = \int_{-\infty}^\infty p(\epsilon)\DELTA_\epsilon d\epsilon,
\eeq
with the operators $\DELTA_\epsilon$ given by $[\DELTA_\epsilon n_t](E)=n_t(E-\epsilon)-n_t(E)$. Using the commutators
\beqar
  [\E,\DELTA_\epsilon] & = & \epsilon\DELTA_\epsilon \\
  {}[\se,\re ] & = & [\m,\re] = 0
\eeqar
we can now collect the operators in one exponential:
\beqar
  \Se\M\R & = & \N e^{-\beta\E} e^{\m} e^{\re} =
  \N e^{-\beta\E} e^{\int p(\epsilon)\DELTA_\epsilon d\epsilon} e^k \\
   & = & \N \exp\left(\int
   \frac{\beta\epsilon}{1-e^{-\beta\epsilon}} p(\epsilon) \DELTA_\epsilon d\epsilon
   -\beta\E\right), \nonumber
\eeqar
which immediately gives the continuum equation
\beq
  \del_t n_t(E) = \int_{-\infty}^{\infty} \hat{p}(\epsilon)
  n_t(E\! -\! \epsilon) d\epsilon + \beta[\hat{\mu}(t)-E] n_t(E)
  \label{LargeKContEq}
\eeq
for $\hat{\mu}(t)$ such that $n_t$ remains normalized, where $\hat{p}(\epsilon) = \frac{\beta\epsilon}{1-e^{-\beta\epsilon}} p(\epsilon)$. Note that we have collected several $E$-independent factors in $\hat{\mu}(t)$, including $\R=K$ and factors from the commutators. The discrete evolution equation only specifies that $n_t$ must be normalized for integer $t$, but we can choose $\hat{\mu}(t)$ such that $n_t$ is normalized for all $t$.

As before, there are classes of models that are equivalent. Note that rescaling time now also rescales temperature, and the mutation rate is rescaled in a nontrivial manner\footnote{Two models are equivalent if and only if $\hat{p}_1/\beta_1=\hat{p}_2/\beta_2$.}. Given Eq.~(\ref{LargeKSelectionApprox}), the mean-field behaviors of ``equivalent" models are indeed identical, and the only effective difference between equivalent models is how often we restrict the population size to $N$. However, since the selection is highly competitive, {\it i.e.}, we remove some DNA that is as good as or better than what remains, that difference is significant.

While the mean-field theory has a formal continuum limit, the evolution process itself is inherently discrete in this regime: Toward the continuum limit, frequent, strong selection will cause a blow-up of the statistical noise from the selection process. This is in contrast to the traditional evolution models, for which the evolution can be described by a continuum process from the start~\cite{TsiLevKes,KesLevRidTsi,RidLevKes}.

\end{document}